\documentclass[aip,jmp,amsmath,amssymb,reprint]{revtex4-1}

\usepackage{empheq}
\usepackage{graphicx}
\usepackage{dcolumn}
\usepackage{bm}

\usepackage{cprotect}
\newcommand*{\citen}[1]{%
  \begingroup
    \romannumeral-`\x 
    \setcitestyle{numbers}%
    \cite{#1}%
  \endgroup
}
\usepackage{soul,color}
\usepackage{etoolbox}
\usepackage{floatrow}
\usepackage{nicefrac}

\begin{document}

\title{High-dimensional dynamics in a single-transistor oscillator containing Feynman-Sierpi\'{n}ski resonators: effect of fractal depth and irregularity}

\author{Ludovico Minati}
\altaffiliation{Author to whom correspondence should be addressed. Electronic addresses: ludovico.minati@ifj.edu.pl, minati.l.aa@m.titech.ac.jp and lminati@ieee.org. Tel.: +39 335 486 670. URL: http://www.lminati.it.}
\affiliation{Complex Systems Theory Department, Institute of Nuclear Physics - Polish Academy of Sciences (IFJ-PAN), 31-342 Krak\'ow, Poland}
\affiliation{Tokyo Tech World Research Hub Initiative, Institute of Innovative Research - Tokyo Institute of Technology, 226-8503 Yokohama, Japan}
\affiliation{Center for Mind/Brain Sciences (CIMeC), University of Trento, 38123 Trento, Italy}
\author{Mattia Frasca}
\affiliation{Department of Electrical Electronic and Computer Engineering (DIEEI), University of Catania, 95131 Catania, Italy}
\author{Gianluca Giustolisi}
\affiliation{Department of Electrical Electronic and Computer Engineering (DIEEI), University of Catania, 95131 Catania, Italy}
\author{Pawe\l\ O\'{s}wi\c{e}cimka}
\affiliation{Complex Systems Theory Department, Institute of Nuclear Physics - Polish Academy of Sciences (IFJ-PAN), 31-342 Krak\'ow, Poland}
\author{Stanis\l aw Dro\.{z}d\.{z}}
\affiliation{Complex Systems Theory Department, Institute of Nuclear Physics - Polish Academy of Sciences (IFJ-PAN), 31-342 Krak\'ow, Poland}
\affiliation{Faculty of Physics, Mathematics and Computer Science, Cracow University of Technology, 31-155 Krak\'ow, Poland}
\author{Leonardo Ricci}
\affiliation{Department of Physics, University of Trento, 38123 Trento, Italy}
\affiliation{Center for Mind/Brain Sciences (CIMeC), University of Trento, 38123 Trento, Italy}

\date{\today}
\begin{abstract}
Fractal structures pervade nature and are receiving increasing engineering attention towards the realization of broadband resonators and antennas. We show that fractal resonators can support the emergence of high-dimensional chaotic dynamics even in the context of an elementary, single-transistor oscillator circuit. Sierpi\'{n}ski gaskets of variable depth are constructed using discrete capacitors and inductors, whose values are scaled according to a simple sequence. It is found that in regular fractals of this kind each iteration effectively adds a conjugate pole/zero pair, yielding gradually more complex and broader frequency responses, which can also be implemented as much smaller Foster equivalent networks. The resonators are instanced in the circuit as one-port devices, replacing the inductors found in the initial version of the oscillator. By means of a highly simplified numerical model, it is shown that increasing the fractal depth elevates the dimension of the chaotic dynamics, leading to high-order hyperchaos. This result is overall confirmed by SPICE simulations and experiments, which however also reveal that the non-ideal behavior of physical components hinders obtaining high-dimensional dynamics. The issue could be practically mitigated by building the Foster equivalent networks rather than the verbatim fractals. Furthermore, it is shown that considerably more complex resonances, and consequently richer dynamics, can be obtained by rendering the fractal resonators irregular through reshuffling the inductors, or even by inserting a limited number of focal imperfections. The present results draw attention to the potential usefulness of fractal resonators for generating high-dimensional chaotic dynamics, and underline the importance of irregularities and component non-idealities.
\end{abstract}
\maketitle

\begin{quotation}
The morphology of diverse natural objects is knowingly self-similar across levels of scale. This feature, referred to as fractality, is observed for example in the shape of coastlines, vegetables, single neurons and even entire brains. Its origin, albeit often ultimately elusive, has at times been ascribed to certain dynamical properties such as operation close to a critical point. On the other hand, comparatively limited attention has been given to the impact that the presence of fractal structure can in itself have on the dynamics of a non-linear system. Since it is well known that fractal patterns may realize complex resonances in a compact and efficient manner, we speculate that they could also be important for supporting the generation of high-dimensional dynamics. Here, this hypothesis is tested through realizing fractal resonators by means of traditional inductors and capacitors, and instancing them into an elementary chaotic oscillator circuit. It is found that increasingly deep fractal resonators yield more complex dynamics, obtaining which is however hindered by non-ideal component behavior. Remarkably, this issue can be alleviated via introducing irregularities, which corrupt the self-similarity and yield vastly richer resonances compared to perfectly regular fractals. Such results lead to speculating about the possible relevance of the truncated, irregular fractal trees of dendrites and axonal outgrowths found in the brain for sustaining its high-dimensional dynamics.
\end{quotation}

\section{INTRODUCTION}\label{intro}
Since the notion of fractal was formalized by Benoit Mandelbrot in year 1975, an extraordinary amount of experimental evidence has accumulated indicating that self-organized systems of the most diverse types tend to generate spatial as well as temporal patterns which recur across levels of scale. Besides popular examples in ecology, such as the shape of clouds, coastlines and some vegetables, some of the most consistent observations arguably come from econophysics and physiology\cite{mandelbrot1983,mandelbrot2004,kwapien2012}. As regards anatomy, the arborization of blood vessels, of the bronchi and, in particular, the morphology of single neurons and their projections all way up to the folding pattern of the brain cortex display fractal properties \cite{caserta1990,alves1996,werner2010,diieva2016}. Perhaps even more remarkably, the virtual totality of physiological signals yield strong signatures of temporal self-similarity related to health and disease states, with well-known examples found in cardiac and cerebral dynamics\cite{havlin1995,goldberger2002}.\\
The reason why fractality is so pervasive in nature ultimately remains elusive. The generation of self-similar patterns appears to be inherent in the complex dynamics of many self-organized systems, such as dwelling close to a critical point in a second-order phase transition\cite{bak1994,kwapien2012}. At the same time, it appears plausible that the presence of self-similar structure in the physical morphology of a system or in the topology of network may impact the dynamics, perhaps considerably. While this topic remains relatively under-investigated, it has been suggested that in the case of the brain, truncated fractals may be fundamentally important, in that they confer an ``intermediate'' level of predictability to the dynamics; excessively shallow fractals would not be capable of supporting sufficient complexity, whereas overly deep ones would lead to effectively unpredictable activity\cite{paar2001}. This perspective may be related to the viewpoint that the brain is principally driven by high-dimensional non-stationary dynamics, which are transiently collapsed down to lower-dimensional dynamics in order to implement coding functions while responding to specific stimuli\cite{singer2016}.\\
Fractals are also remarkable in that, in the limit of infinite number of iterations (depth), they collapse an unlimited perimeter into a bounded area, and an unlimited area into a bounded volume. Similarly, truncated fractals practically realize very high surface-to-volume ratios. The geometric properties of fractals are increasingly relevant to engineering, for example in the fields of optical transmission \cite{bao2007}, quantum interference \cite{carlier2004} and fractal electronics \cite{fairbanks2011}. The latter area encompasses heterogeneous approaches such as fractal-based layouts, self-assembly techniques leading to fractal structures, and fractal circuits realized with traditional discrete components.\\
In particular, fractal-based layouts have shown considerable promise in the miniaturization of devices, such as antennas, for realizing which it is crucial to fit long wavelengths in small areas and volumes\cite{baliarda2000,cohen2015}, and even in stretchable electronics, where fractal design concepts enable implementing advanced functions through exotic mechanical properties\cite{fan2014}. Self-assembly techniques, and more specifically diffusion-limited aggregation, have furthermore been used to obtain fractal geometries yielding non-linear conduction behaviors, for example in antimony aggregates on graphite, which find applications in high-sensitivity sensors \cite{scott2006,fairbanks2011}. Finally, fractal circuits are easily obtained through realizing the branches of geometric fractals with discrete components or combinations thereof. Some of the circuits thus obtained have peculiar, apparently paradoxical properties in that, in the limit of infinite number of iterations, they enable power dissipation through purely reactive components\cite{chen2017,alonso2017}. Fractal ring and cell topologies also have practical importance in the realization of high-frequency oscillators on integrated circuits\cite{kim2015,choi2015}.\\
In this work, we address the overarching question of whether fractal structures could have a role in enabling and supporting the emergence of high-dimensional chaotic dynamics, e.g. having $D>3$. We do so in the context of an elementary, autonomous single-transistor oscillator, which is the smallest member of a large family of atypical chaotic circuits that were recently obtained by means of a heuristically-driven random search procedure \cite{minati2017}. We focus on a specific fractal structure, i.e., the \emph{Feynman-Sierpi\'{n}ski ladder} or \emph{resonator}, which has been introduced in Ref. \citen{chen2017} starting from the well-known Sierpi\'{n}ski gasket\cite{mandelbrot1983} and embodying its branches by means of inductors and capacitors. In the present context, such resonator is treated as a one-port element and instanced as a replacement for each of two inductors present in the initial oscillator.\\
In Section \ref{fractal}, analytic procedures for calculating the impedance of the fractal resonator and for obtaining simpler equivalent networks are introduced. In Section \ref{oscillator}, the oscillator circuit and a simplified numerical model are presented, followed by results from simulations. In Section \ref{experimental}, experimental measurements of the resonator frequency responses and oscillator time-series are reported, separately for regular and irregular realizations of the fractals. Finally, in Section \ref{discussion} the results are discussed from the viewpoint of their relevance to electronic chaotic oscillators, as well as more generally.

\begin{figure*}
\centering
\includegraphics[width=0.9\textwidth]{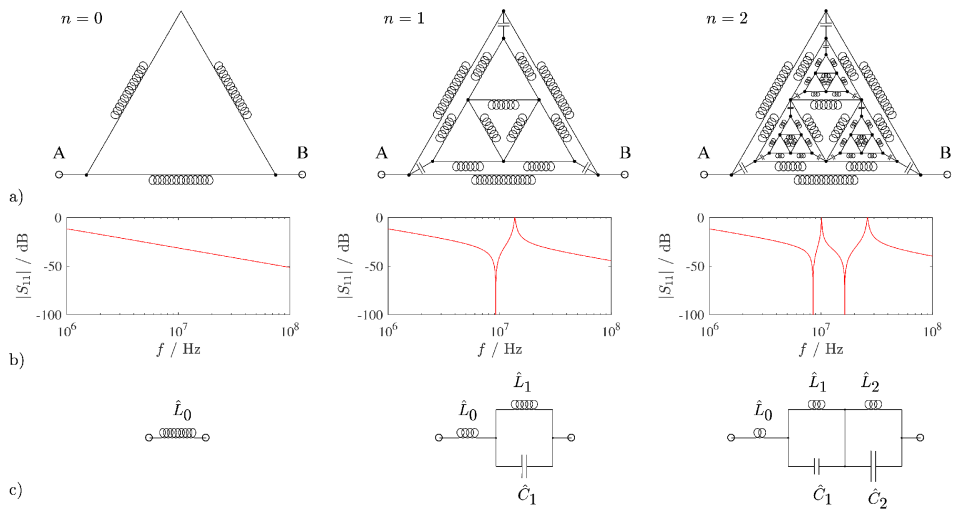}
\cprotect\caption{Feynman-Sierpi\'{n}ski fractal resonator at $n=0,1,2$ iterations. a) Network topology, denoting the two nodes A, B available for external connection as a one-port device. b) Frequency response expressed as reflection coefficient $S_{11}$. c) Equivalent networks yielded by the Foster method. Number of turns in inductor symbols is proportional to inductance.\label{fig:fig1}}
\end{figure*}
\begin{figure*}
\centering
\includegraphics[width=0.8\textwidth]{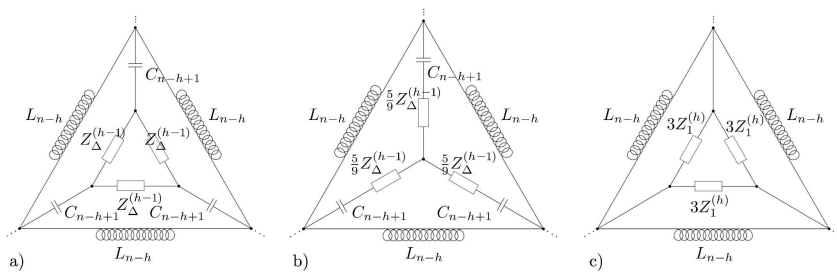}
\cprotect\caption{Recursive steps for obtaining analytically the complex impedance of the one-port resonator device $Z_\textrm{AB}^{(n)}$ at iteration $n$. a) Initial $\Delta$ configuration equivalent to the fractal resonator. b) Intermediate circuit obtained after transformation from the $\Delta$ to the Y configuration. c) Final circuit obtained after transformation back to the $\Delta$ configuration.\label{fig:fig2}}
\end{figure*}
\begin{figure}
\centering
\includegraphics[width=0.95\textwidth]{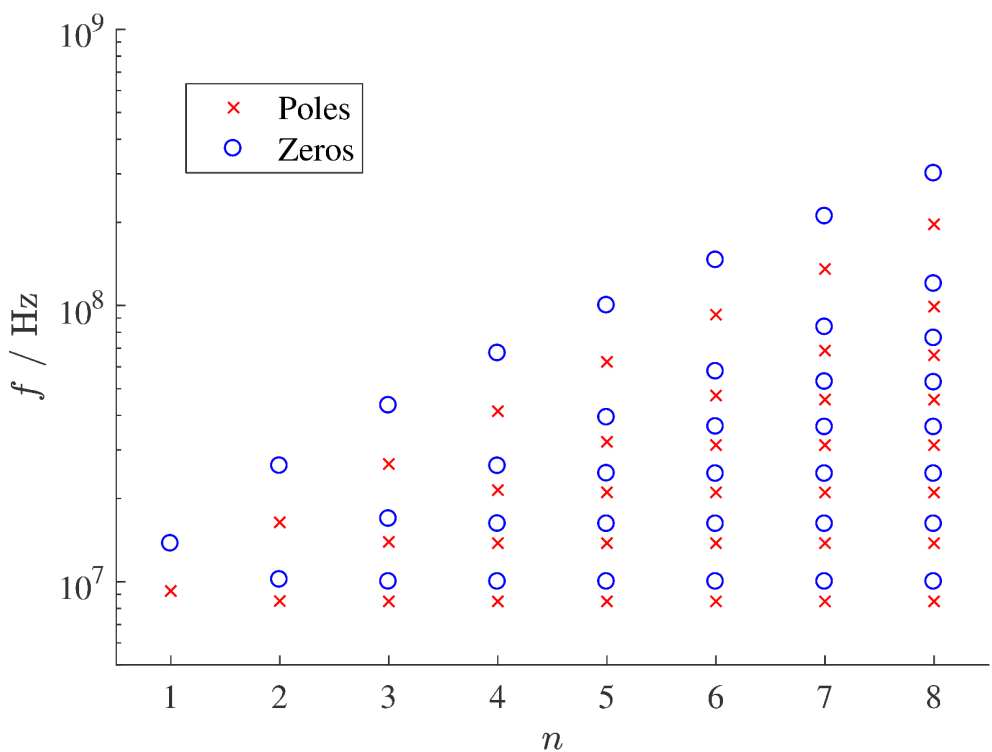}
\cprotect\caption{Frequency distribution of non-trivial poles and zeros of $Z_{AB}^{(n)}$ as a function of fractal depth $n$.\label{fig:fig3}}
\end{figure}
\section{FEYNMAN-SIERPI\'{N}SKI FRACTAL RESONATOR}\label{fractal}
\subsection{Construction}\label{fractal_construction}
Feynman-Sierpi\'nski fractal resonators were constructed according to the following procedure, which represents an adaptation of the canonical construction of the Sierpi\'nski gasket inspired by Feynman's infinite LC ladder. This procedure, alongside fundamental theorems about the symmetry and impedance of fractal LC networks, has been introduced in Refs. \citen{chen2017,alonso2017}.\\
First, three inductors are interconnected to form a triangle, which represents iteration step $n=0$. Second, three capacitors are connected between its vertices and three pairs of series inductors, forming a second triangle inscribed within the first one. Third, the mid-points of the inductor pairs are interconnected between themselves via three more inductors, yielding the fractal at iteration step $n=1$. As represented in Fig. \ref{fig:fig1}a, further iterations are realized by arbitrarily repeating the second and third steps, realizing a clearly self-similar topology. At iteration $n$, the fractal circuit contains $\sum_{i=0}^n3^{i+1}$ inductors and $\sum_{i=1}^n3^i$ capacitors.\\
While previous work has addressed the paradox of power dissipation\cite{chen2017,alonso2017} in the asymptote $n\rightarrow\infty$, here we are concerned with the behavior at small values of $n$, that is, with truncated Feynman-Sierpi\'nski fractals. To the authors' knowledge their resonances have not yet been systematically investigated, and our interest is motivated by the fact that they could be relevant for supporting complex dynamics in non-linear dynamical systems.\\
Reflecting the basic property of the Sierpi\'nski gasket that side lengths are halved at each iteration, the value of the inductors was halved at each level, i.e. $L_{i+1}=L_i/2$. As the capacitors, albeit necessary to couple the different levels of the fractal, do not directly correspond to a geometric feature of the Sierpi\'nski gasket, their value $C_i=C$ was kept constant across iterations. These choices are ultimately arbitrary and different arguments could be offered in support of other inductance and capacitance sequences, e.g. $L_i=L$, $C_{i+1}=C_i/2$ etc. A comprehensive study of such cases is beyond the scope of the present work.\\
As detailed in Section \ref{oscillator}, the fractal resonators were inserted in the initial oscillator circuit according to a configuration wherein they are treated as one-port devices, i.e. one vertex of the outer triangle is not externally connected. To facilitate comparison between the analytical calculations and the experimental data given in Section \ref{experimental}, the frequency responses are accordingly charted in terms of the reflection coefficient $S_{11}$, corresponding to impedance $Z_\textrm{AB}=Z_\textrm{o}(1+S_{11})\big/(1-S_{11})$, where $Z_\textrm{o}=50\ \Omega$. While at step $n=0$ the device is equivalent to an inductor, i.e. has no self-resonance, as shown in Fig. \ref{fig:fig1}b each iteration has the effect of adding one pair of conjugate imaginary poles and one pair of conjugate imaginary zeros. In order to realize resonances suitable for excitation by the chosen physical oscillator detailed in Section \ref{oscillator}, namely in the range 10-100 MHz, we set $L_0=22\ \mu\textrm{H}$ and $C=22\textrm{ pF}$.

\subsection{Impedance calculation}\label{fractal_impedance}
The impedance $Z_\textrm{AB}^{(n)}$ of the one-port device representing the $n$-th iteration of the fractal can be obtained analytically as follows.\\
First, let us consider the $\Delta$ configuration equivalent to the Feynman-Sierpi\'nski resonator at iteration $1\leq h\leq n$ (Fig. \ref{fig:fig2}a), and denote the impedance of each branch as $Z_{\Delta}^{(h)}$; this impedance is calculated iteratively. Second, assuming that $Z_{\Delta}^{(h-1)}$ is known, the circuit is transformed from the $\Delta$ to the equivalent $Y$ configuration (Fig. \ref{fig:fig2}b), then $Z_{1}^{(h)}$ can be trivially calculated as two impedances in series, i.e., 
\begin{equation}
\label{eq:Zeq1ite_a}
\begin{array}{l}
Z_{1}^{(h)}=\frac{5}{9}Z_{\Delta}^{(h-1)}+\frac{1}{sC_{n-h+1}}\textrm{ .}
\end{array}
\end{equation}
Third, the circuit is transformed back from the $Y$ to the $\Delta$ configuration (Fig. \ref{fig:fig2}c), and $Z_{\Delta}^{(h)}$ is obtained as two impedances in parallel, i.e.,
\begin{equation}
\label{eq:Zeq1ite_b}
\begin{array}{l}
Z_{\Delta}^{(h)}=\frac{3Z_{1}^{(h)}sL_{n-h}}{3Z_{1}^{(h)}+sL_{n-h}}\textrm{ .}
\end{array}
\end{equation}
After iteration over $h=1,\ldots,n$ has been completed, i.e., $h=n$, one has 
\begin{equation}
\label{eq:Zeq1ite_c}
Z_\textrm{AB}^{(n)}=\frac{2}{3}Z_{\Delta}^{(n)}\textrm{ .}
\end{equation}
For the chosen sequences $L_n$ and $C_n$, each iteration adds one pair of conjugate imaginary poles and one pair of conjugate imaginary zeros at a higher frequency which, with increasing $n$, shift downwards realizing a distribution that approximates a log-periodic pattern (Fig. \ref{fig:fig3}).\\
Explicit expressions are given for low values of $n$: for $n=1$, one has
\begin{equation}
\label{eq:Zeq1itek1}
Z_\textrm{AB}^{(1)}=\frac{2}{3}\frac{sL_0\left ( 1+\frac{5}{9}s^2L_1C_1\right)}{1+s^2C_1\left ( \frac{1}{3}L_0+\frac{5}{9}L_1\right)}
\end{equation}
\noindent and for $n=2$, one has
\begin{equation}
\label{eq:Zeq2itek2}
Z_\textrm{AB}^{(2)}=\frac{2}{3}\frac{\parbox{2.7in}{$sL_0\big ( \frac{25}{81}s^4L_1C_1L_2C_2+$\\\hspace*{0.5cm}$s^2\big( \frac{1}{3}C_2L_1+\frac{5}{9}C_2L_2+\frac{5}{9}C_1L_1\big) +1 \big)$}} {\parbox{2.7in}{$s^4\big(  \frac{25}{81}L_1C_1L_2C_2+L_0C_1\big( \frac{1}{9}C_2L_1+\frac{5}{27}C_2L_2 \big)\big)+$\\\hspace*{0.5cm}$s^2\big( \frac{1}{3}C_2L_1+\frac{5}{9}C_2L_2 +$\\\hspace*{0.5cm}$\frac{5}{9}C_1L_1+\frac{1}{3}C_1L_0\big)+1$}}\textrm{ .}
\end{equation}
\subsection{Equivalent network}\label{fractal_foster}
As the resonator includes only energy-storing components, its impedance is a lossless positive-real transfer function and all poles and zeros lie on the imaginary axis and alternate. It is therefore possible to derive a series LC network having impedance equal to that of the fractal circuit using the Foster method\cite{kuo2006}. To this end, we first consider the partial fraction expansion of $Z_\textrm{AB}^{(n)}$; as the system always has a zero at the origin $s=0$, it reads
\begin{equation}
\label{eq:ZeqFoster}
Z_\textrm{AB}^{(n)}=K_0s+\frac{K_1s}{s^2+\omega_1^2}+\ldots+\frac{K_ns}{s^2+\omega_n^2}
\end{equation}
\noindent where $\omega_1,\ldots,\omega_n$ are the pole frequencies, and $K_0,\ldots,K_n$ are constants which can be determined by equating Eq. (\ref{eq:Zeq1ite_c}) and Eq. (\ref{eq:ZeqFoster}), and applying the principle of polynomial identity. For succinctness, the superscript $(n)$ is omitted and left implicit in the notation for these parameters and those derived from them. The impedance $Z_\textrm{AB}^{(n)}$ is represented as elementary impedances in series (Fig. \ref{fig:fig1}c), where the first block contains an inductor only, here labeled as $\hat{L}_0$, and all remaining blocks with $j=1,\ldots,n$ each consist of a capacitor $\hat{C}_j$ and an inductor $\hat{L}_j$ connected in parallel.\\
For a Feynman-Sierpi\'nski resonator at iteration $n$, there are $n+1$ blocks and $2n+1$ components in the equivalent Foster network. To perform the synthesis, the network elements are associated with the terms in the partial fraction expansion as follows
\begin{equation}
\label{eq:ZeqFosterEquivalence1}
\hat{L}_0=K_0
\end{equation}
\noindent and
\begin{equation}
\label{eq:ZeqFosterEquivalence2}
\begin{array}{l}
\hat{C}_j=1/K_j\\
\hat{L}_j=K_j/\omega_i^2
\end{array}
\end{equation}
\noindent for $j=1,\ldots,n$.
This result is particularly relevant to the present purpose, as it enables investigating the effect of increasing fractal depth on dynamics, without incurring the exponentially-increasing number of components required to realize verbatim each level of the fractal.\\
Considering as an example $n=1$, $Z_\textrm{AB}^{(1)}$ is given by Eq. (\ref{eq:Zeq1itek1}) and its partial fraction expansion yields the following coefficients
\begin{equation}
\label{eq:ZeqFosterk1}
\begin{array}{l}
K_0=\frac{\frac{10}{27}L_0L_1}{\frac{1}{3}L_0+\frac{5}{9}L_1}\\
\omega_1^2=\left(C_1\left(\frac{1}{3}L_0+\frac{5}{9}L_1\right)\right)^{-1}\\
K_1=K_0\left(\frac{1}{\frac{5}{9}L_1C_1}-\omega_1^2\right)
\end{array}
\end{equation}
\noindent from which 
\begin{equation}
\label{eq:ZeqFosterEquivalence1k1}
\begin{array}{l}
\hat{L}_0=\frac{\frac{10}{27}L_0L_1}{\frac{1}{3}L_0+\frac{5}{9}L_1}\\
\hat{C}_1=\left (K_0\left(\frac{1}{\frac{5}{9}L_1C_1}-\omega_1^2\right)\right )^{-1}\\
\hat{L}_1=K_0C_1\left(\frac{1}{\frac{5}{9}L_1C_1}-\omega_1^2\right)\left(\frac{1}{3}L_0+\frac{5}{9}L_1\right)
\end{array}
\end{equation}
For higher numbers of iterations, closed-form expressions become prohibitively long and numerical solution is preferable. The inductance and capacitance values realizing the equivalent networks for $n=0,\ldots,3$ are given in Table \ref{tab:tabella1}, and used in subsequent numerical simulations.\\

\begin{table}
\caption{Component values realizing the equivalent LC networks for the fractal resonators having depth $n=0,\ldots,3$. Inductances $\hat{L}$ and capacitances $\hat{C}$ are expressed in $\mu\textrm{H}$ and $\textrm{pF}$ respectively.}\label{tab:tabella1}
\begin{tabular}{cccccccc}
\hline
Depth $n$ & $\hat{L}_0$ & $\hat{L}_1$ & $\hat{C}_1$ & $\hat{L}_2$ & $\hat{C}_2$ & $\hat{L}_3$ & $\hat{C}_3$\\\hline
0 & $14.67$ & $\emptyset$ & $\emptyset$ & $\emptyset$ & $\emptyset$ & $\emptyset$ & $\emptyset$\\
1 & $6.67$ & $8.00$ & $36.97$ & $\emptyset$ & $\emptyset$ & $\emptyset$ & $\emptyset$\\
2 & $4.03$ & $5.29$ & $17.78$ & $5.34$ & $65.41$ & $\emptyset$ & $\emptyset$\\
3 & $2.73$ & $3.55$ & $10.02$ & $3.20$ & $40.73$ & $5.18$ & $67.85$\\\hline
\end{tabular}
\end{table}

\begin{figure*}
\centering
\includegraphics[width=0.87\textwidth]{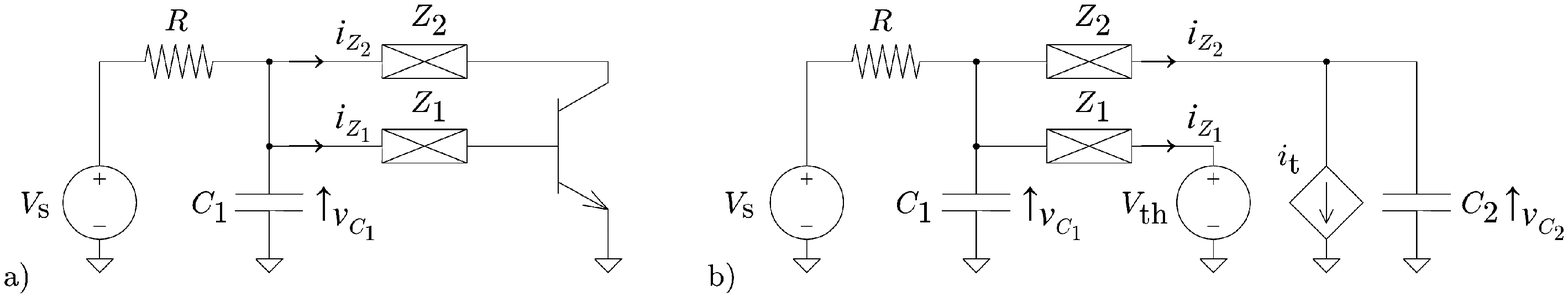}
\cprotect\caption{Single-transistor chaotic oscillator. a) Physically-realized circuit derived from Ref. \citen{minati2017}. The elements $Z_1$ and $Z_2$ denote separate instances of the fractal resonator, shown in its verbatim form in Fig. \ref{fig:fig1}a and in its equivalent form in Fig. \ref{fig:fig1}c. b) Simplified model, corresponding to Eqs. (\ref{eq:simploscmodel2}) and (\ref{eq:simploscmodel2b}). See Section \ref{oscillator} for detailed description.\label{fig:fig4}}
\end{figure*}

\section{SINGLE-TRANSISTOR OSCILLATOR}\label{oscillator}
\subsection{Initial circuit and its simplified model}\label{oscillator_model}
To investigate the possible effect of increasing fractal depth on dynamics, we inserted the fractal resonators in an elementary bipolar-junction transistor oscillator, replacing the initial inductors with more general impedances $Z_1$ and $Z_2$ (Fig. \ref{fig:fig4}a). The chosen circuit was introduced in Ref. \citen{minati2017}, wherein a large-scale random search recurrently identified it as the smallest topology giving rise to chaotic behavior. It is an autonomous oscillator powered by a fixed voltage source $V_\textrm{s}$ connected via a resistor whose value $R$ is treated as control parameter, as a function of which diverse dynamics including spiral, phase-coherent attractors, attractors resembling the R{\"o}ssler funnel attractor and spiking attractors can be observed.\\
The circuit consists of a single NPN transistor in common-emitter configuration, whose base and collector terminals are connected to the resistor and to a capacitor $C_1$ (distinct from that in Section \ref{fractal}) via two separate impedances $Z_1$ and $Z_2$, which in the simplest implementation ($n=0$) correspond to two inductors. In Ref. \citen{minati2017}, the behavior of this circuit was studied as a function of the values of the inductors and capacitor; here, we instead investigate its behavior as the initial inductors (corresponding to the case $n=0$ in Fig. \ref{fig:fig1}a) are replaced with fractal resonators of varying depth $n>0$, maintaining $Z_1=Z_2$ unless otherwise indicated.\\
To aid understanding and numerical simulation, it is possible to considerably simplify the transistor model via three assumptions\cite{millman1987} (Fig. \ref{fig:fig4}b). First, since during oscillation the base-emitter voltage $v_\textrm{BE}$ remains approximately constant, a fixed voltage source $V_\textrm{th}\approx v_\textrm{BE}$ is connected between $Z_1$ (base terminal) and ground. Second, the junction capacitances are collapsed into a single capacitor $C_2$  (distinct from that in Section \ref{fractal}) connected between $Z_2$ (collector terminal) and ground; despite having a relatively small value, this capacitor is essential for sustaining oscillation (details not shown). Third, non-linear amplification is represented by a controlled current sink connected to $Z_2$ (collector terminal), whose intensity $i_\textrm{t}$ depends in a stylized manner on the base current $i_{Z_1}$ and on the collector voltage $v_{C_2}$.\\
For the initial circuit wherein $n=0$ and $Z_{k=1,2}=s \hat{L}_0^{(k)}$, one can thus take as state variables the voltage drops across the two capacitors, i.e. $v_{C_1}$ and $v_{C_2}$, and the currents through the two inductors, i.e. $i_{\hat{L}_0^{(1)}}$ and $i_{\hat{L}_0^{(2)}}$. It should be clarified that in this context the superscript $(k=1,2)$ denotes the instance of the fractal, rather than its depth as $(n)$ does in Section \ref{fractal}. Applying Kirchhoff's laws, the state equations are obtained
\begin{empheq}[left=\empheqlbrace]{align}
\label{eq:simploscmodel1}
\frac{\textrm{d}v_{C_1}}{\textrm{d} t}&=\frac{V_\textrm{s}-v_{C_1}}{RC_1}-\frac{i_{\hat{L}_0^{(1)}}+i_{\hat{L}_0^{(2)}}}{C_1}\nonumber\\
\frac{\textrm{d}v_{C_2}}{\textrm{d} t}&=\frac{i_{\hat{L}_0^{(2)}}-i_\textrm{t}}{C_2}\nonumber\\
\frac{\textrm{d}i_{\hat{L}_0^{(1)}}}{\textrm{d} t}&=\frac{v_{C_1}-V_\textrm{th}}{\hat{L}_0^{(1)}}\\
\frac{\textrm{d}i_{\hat{L}_0^{(2)}}}{\textrm{d} t}&=\frac{v_{C_1}-v_{C_2}}{\hat{L}_0^{(2)}}\nonumber\textrm{ ,}
\end{empheq}
wherein one can for example set $i_\textrm{t}=\alpha\left(i_{\hat{L}_0^{(1)}},v_{C_2}\right)$, with
\begin{equation} 
\label{eq:simploscmodel1b}
\alpha\left(x,y\right)=\beta \Gamma\left(x\right) \tanh\left(y/2V_{\textrm{th}}\right)\textrm{ ,}
\end{equation}
and $\Gamma(x)=x$ for $x>0$ and $\Gamma(x)=0$ for $x \leq 0$, $\beta$ denotes the forward current gain, $\Gamma(x)$ prohibits the amplification of negative base currents, which is important for convergence, and $\tanh(y)$ conveniently implements a non-linear amplification effect; here, its argument was empirically scaled by $V_{\textrm{th}}$, but similar spectra and sections are obtained for $V_{\textrm{th}}\rightarrow0$ in this equation, which approximates a step function. In general, the form and parameters of Eq. (\ref{eq:simploscmodel1b}) were chosen arbitrarily: they only loosely map onto transistor equations and are not critical, moreover as shown in Subsection \ref{oscillator_spice} analogous results are obtained with a more realistic transistor model.\\
For the more general case $n>0$ where the impedances $Z_k$ rather than inductors represent the resonant networks, the state equations can be extended as 
\begin{empheq}[left=\empheqlbrace]{align}
\label{eq:simploscmodel2}
\frac{\textrm{d}v_{C_1}}{\textrm{d}t}&=\frac{V_{\textrm{s}}-v_{{C_1}}}{RC_1}-\frac{i_{\hat{L}_0^{(1)}}+i_{\hat{L}_0^{(2)}}}{C_1}\nonumber\\
\frac{\textrm{d}v_{C_2}}{\textrm{d}t}&=\frac{i_{\hat{L}_0^{(2)}}-i_\textrm{t}}{C_2}\nonumber\\
\frac{\textrm{d}i_{\hat{L}_0^{(1)}}}{\textrm{d}t}&=\frac{v_{C_1}-\sum{v_{\hat{C}^{(1)}_j}}-V_{\textrm{th}}}{\hat{L}_0^{(1)}}\\
\frac{\textrm{d}i_{\hat{L}_0^{(2)}}}{\textrm{d}t}&=\frac{v_{C_1}-\sum{v_{\hat{C}^{(2)}_j}}-v_{C_2}}{\hat{L}_0^{(2)}}\nonumber
\end{empheq}
where in addition one has, for each of the two instances $k=1,2$ and level $j=1,\ldots,n$ of the fractal, 
\begin{empheq}[left=\empheqlbrace]{align}
\label{eq:simploscmodel2b}
\frac{\textrm{d}v_{\hat{C}^{(k)}_j}}{\textrm{d}t}&=\frac{i_{\hat{L}_0^{(k)}}-i_{\hat{L}^{(k)}_j}}{\hat{C}^{(k)}_j}\nonumber\\
\frac{\textrm{d}i_{\hat{L}^{(k)}_j}}{\textrm{d}t}&=\frac{v_{\hat{C}^{(k)}_j}}{\hat{L}^{(k)}_j}\textrm{ .}
\end{empheq}
In other words, for $n$ iterations of the fractal, a dynamical system of order $4+4n$ is obtained. We underline that when $n>0$ the inductors $\hat{L}_0^{(k)}$ present in the initial circuit correspond to the first block in the Foster networks (Fig. \ref{fig:fig1}c). Representing some characteristics of the physical oscillator detailed in Ref. \citen{minati2017} and Section \ref{experimental}, we set $V_{\textrm{s}}=2.5\textrm{ V}$, $C_1=270\textrm{ pF}$, $C_2=5\textrm{ pF}$, $V_{\textrm{th}}=0.6\textrm{ V}$, $\beta=200$ and swept the control parameter $R\in[0,2000]\ \Omega$.\\

\begin{table}
\caption{Time constants $\tau^{(k)}_j$ and $T^{(k)}_j$ representing the capacitances and inductances as a function of fractal depth $n$, wherein $j\le n$, $k=1,2$, and given $R=1000\ \Omega$. All values are expressed in ns.}\label{tab:tabella2}
\begin{tabular}{cccccccc}
\hline
Depth $n$ & $T^{(k)}_0$ & $T^{(k)}_1$ & $T^{(k)}_2$ & $T^{(k)}_3$ & $\tau^{(k)}_1$ & $\tau^{(k)}_2$ & $\tau^{(k)}_3$\\\hline
0 & 14.67 & $\emptyset$ & $\emptyset$ & $\emptyset$ & $\emptyset$ & $\emptyset$ & $\emptyset$\\
1 & 6.67 & 8.00 & $\emptyset$ & $\emptyset$ & 36.97 & $\emptyset$ & $\emptyset$\\
2 & 4.03 & 5.29 & 5.34 & $\emptyset$ & 17.78 & 65.41 & $\emptyset$\\
3 & 2.73 & 3.55 & 3.20 & 5.18 & 10.02 & 40.73 & 67.85\\\hline
\end{tabular}
\end{table}

\begin{table*}
\centering
\caption{Kaplan-Yorke dimension $D_\textrm{KY}$ and positive Lyapunov exponents $\lambda_i>0$, estimated via integration of the simplified model (Fig. \ref{fig:fig4}b, Eqs. \ref{eq:rescoscmodel1} and \ref{eq:rescoscmodel2}), as a function of fractal depth $n$. All Lyapunov exponents expressed in units of $\mu\textrm{s}^{-1}$; negative exponents not shown. All values given as mean$\pm$standard deviation. Superscripts $(\approx)$ and $(\wr)$ denote, respectively, inconsistent sign across simulation runs and presence of numerical instability. For $n=0$, $\delta D_\textrm{KY}\approx10^{-4}$ and $\delta \lambda_1\approx10^{-5}$.}\label{tab:tabella3}
\begin{tabular}{ccccccc}
\hline
Depth $n$ & $D_\textrm{KY}$ & $\lambda_1$ & $\lambda_2$ & $\lambda_3$ & $\lambda_4$ & $\lambda_5$\\
\hline
\multicolumn{7}{c}{$R=300\ \Omega$}\\
\hline
0 & $1.1\pm0.0$ & $0.02\pm0.00^{(\wr)}$\\
1 & $5.3\pm0.1$ & $2.70\pm0.12$ & $1.13\pm0.13$ & $0.01\pm0.01^{(\approx)}$\\
2 & $7.2\pm0.2$ & $2.21\pm0.26$ & $0.88\pm0.16$ & $0.17\pm0.10$ & $0.05\pm0.02$\\
3 & $9.0\pm0.2$ & $1.85\pm0.25$ & $0.87\pm0.10$ & $0.32\pm0.09$ & $0.07\pm0.03$ & $0.01\pm0.06^{(\approx)}$\\
\hline
\multicolumn{7}{c}{$R=1500\ \Omega$}\\
\hline
0 & $1.3\pm0.0$ & $0.07\pm0.00$\\
1 & $3.9\pm0.1$ & $0.61\pm0.09$ & $0.03\pm0.03$\\
2 & $8.0\pm0.2$ & $2.26\pm0.20$ & $0.75\pm0.13$ & $0.15\pm0.08$ & $0.05\pm0.02$\\
3 & $10.1\pm0.2$ & $1.64\pm0.12$ & $0.75\pm0.09$ & $0.28\pm0.07$ & $0.11\pm0.03$ & $0.07\pm0.01$\\
\hline
\end{tabular}
\end{table*}

\begin{figure*}
\centering
\includegraphics[width=1\textwidth]{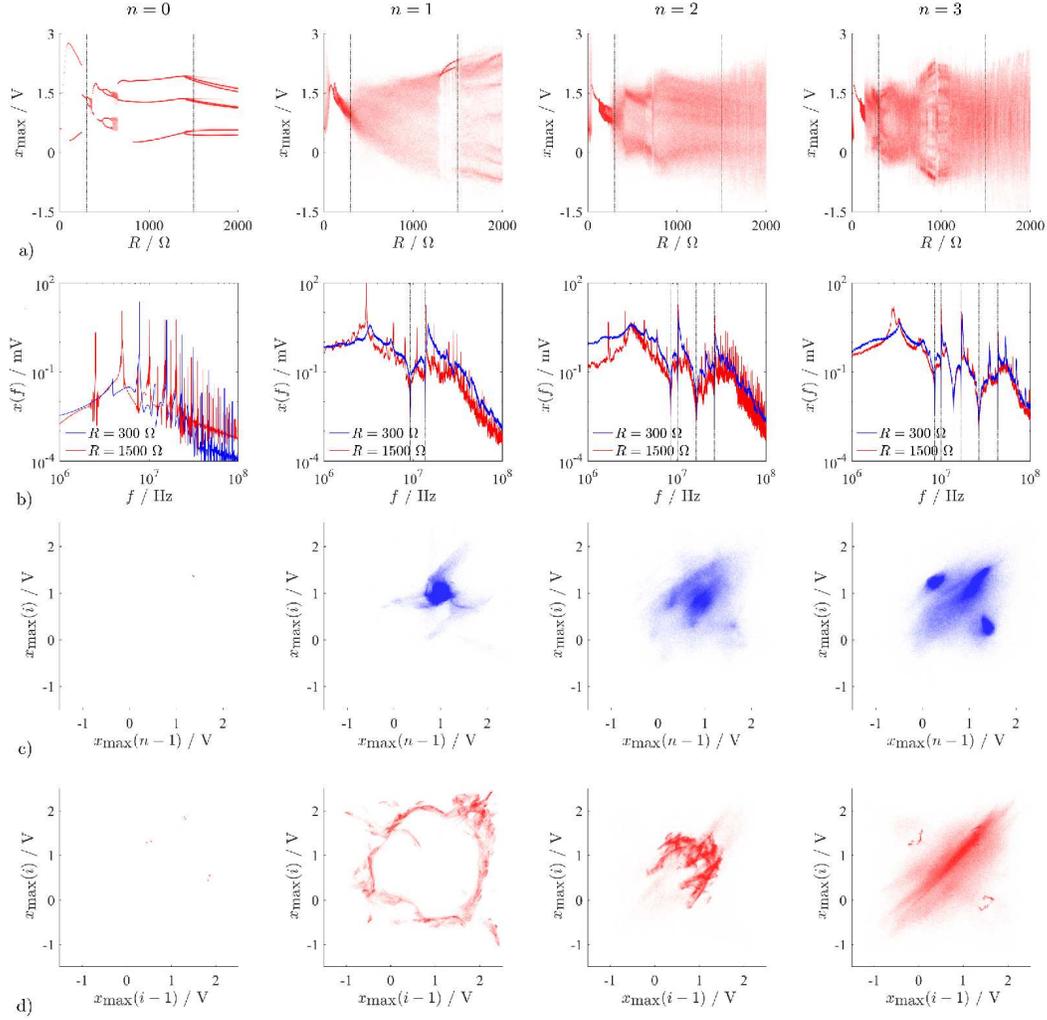}
\cprotect\caption{Numerical simulations of the simplified model (Fig. \ref{fig:fig4}b, Eqs. \ref{eq:rescoscmodel1} and \ref{eq:rescoscmodel2}) at $n=0,1,2,3$ iterations of the fractal. a) Bifurcation diagrams, representing the distribution of local maxima. b) Frequency spectra for $R=300\ \Omega$ and $R=1500\ \Omega$, smoothed for visualization. Dashed lines denote the location of poles and zeros of the fractal resonator impedance $Z_\textrm{AB}$ (Fig. \ref{fig:fig3}). c) and d) Representative zero temporal-derivative Poincar\'{e} sections, respectively for $R=300\ \Omega$ and $R=1500\ \Omega$.\label{fig:fig5}}
\end{figure*}

\subsection{Numerical simulations}\label{oscillator_numerical}
For numerical simulations and analysis it is appropriate to rewrite Eqs. (\ref{eq:simploscmodel2}) and (\ref{eq:simploscmodel2b}) only in terms of voltages, replacing the state variables with $x=v_{C_1}$, $y=v_{C_2}$, $z=Ri_{\hat{L}_0^{(1)}}$ and $w=Ri_{\hat{L}_0^{(2)}}$, and introducing new parameters representing the time constants associated with the inductances and capacitances $\tau_1=RC_1$, $\tau_2=RC_2$, $T^{(1)}_0=\hat{L}_0^{(1)}/R$ and $T^{(2)}_0=\hat{L}_0^{(2)}/R$. One has
\begin{empheq}[left=\empheqlbrace]{align}
\label{eq:rescoscmodel1}
\frac{\textrm{d}x}{\textrm{d}t}&=\frac{V_{\textrm{s}}-x-z-w}{\tau_1}\nonumber\\
\frac{\textrm{d}y}{\textrm{d}t}&=\frac{w-\alpha(z,y)}{\tau_2}\nonumber\\
\frac{\textrm{d}z}{\textrm{d}t}&=\frac{1}{T^{(1)}_0}\left(x-\sum v^{(1)}_j-V_{\textrm{th}} \right)\\
\frac{\textrm{d}w}{\textrm{d}t}&=\frac{1}{T^{(2)}_0}\left(x-\sum v^{(2)}_j-y\right)\nonumber
\end{empheq}
with, for each level $j=1,\ldots,n$ of the fractal, 
\begin{empheq}[left=\empheqlbrace]{align}
\label{eq:rescoscmodel2}
\frac{\textrm{d}v^{(1)}_j}{\textrm{d}t}&=\frac{z-u^{(1)}_j}{\tau^{(1)}_j}\nonumber\\
\frac{\textrm{d}v^{(2)}_j}{\textrm{d}t}&=\frac{w-u^{(2)}_j}{\tau^{(2)}_j}\nonumber\\
\frac{\textrm{d}u^{(1)}_j}{\textrm{d}t}&=\frac{v^{(1)}_j}{T^{(1)}_j}\textrm{ .}\\
\frac{\textrm{d}u^{(2)}_j}{\textrm{d}t}&=\frac{v^{(2)}_j}{T^{(2)}_j}\nonumber
\end{empheq}
In these latter equations, the state variables are similarly replaced with the voltages $v^{(k)}_j=v_{\hat{C}^{(k)}_j}$ and $u^{(k)}_j=Ri_{\hat{L}^{(k)}_j}$, and the inductances and capacitances are replaced by the time constants $\tau^{(k)}_j=R\hat{C}^{(k)}_j$ and $T^{(k)}_j=\hat{L}^{(k)}_j/R$.\\
It is noteworthy that with increasing values of the control parameter $R$, the time constants associated to the capacitances and inductances respectively diverge and vanish in $\lim_{R\to\infty}\tau=\infty$ and $\lim_{R\to\infty}T=0$. Given an intermediate setting of the control parameter $R=1000\ \Omega$, one has $\tau_1=270\textrm{ ns}$ and $\tau_2=5\textrm{ ns}$; these time constants are independent of the fractal depth $n$. On the other hand, the time constants $\tau^{(k)}_j$ and $T^{(k)}_j$, which are identical for $k=1,2$, are given in Table \ref{tab:tabella2} and lie in the range 1-100 ns, all change as a function of the fractal depth $n$, so that deeper fractals render a greater span of time-scales available to the system. This is directly reflected by the distributions of poles and zeros shown in Fig. \ref{fig:fig3}.\\
The rescaled system in Eqs. (\ref{eq:rescoscmodel1}) and (\ref{eq:rescoscmodel2}) was integrated by means of the explicit embedded Runge-Kutta Prince-Dormand $(8,9)$ method\cite{rungekutta}, furthermore implementing the so-called standard method\cite{Benettin_1_1980,Benettin_2_1980,Skokos2010}, which allows evaluating the set of characteristic Lyapunov exponents of a flow directly from its analytic description. While the solver has variable step-size, the state variables were recorded every $\textrm{d}t=0.1~\textrm{ns}$, which is about one order of magnitude smaller than the typical time-scales of the system's evolution, until $t=100\ \mu\textrm{s}$. The Gram-Schmidt orthogonalization process inherent in the standard method was set to take place every $10^3$ steps. Even though their study was outside the scope of the present work, we note that, particularly for large values of $R$, metastable behaviors could become visible in long simulation runs.\\
Based on the sorted Lyapunov exponents $\lambda_i\ge\lambda_{i+1}$ provided by the standard method, the Kaplan-Yorke (or Lyapunov) dimension $D_\textrm{KY}$ was calculated as
\begin{equation}
\label{eq:KaplanYorkeDimension}
{D}_\textrm{KY} = k + \sum_{i=1}^{k}\frac{\lambda_i}{\left|\lambda_{k+1}\right|}\textrm{ ,}
\end{equation}
where $k$ is the maximum integer such that the sum of the $k$ largest exponents is non-negative\cite{KaplanYorke1979}. Integration was repeated 10 times with random initial conditions, and the uncertainty $\delta D_\textrm{KY}$ was estimated as the corresponding standard deviation\cite{FranchiRicci2014}.\\
The bifurcation diagrams revealed that, even though narrow chaotic regions were already apparent at $n=0$ (no fractal iterations, corresponding to the initial circuit; e.g., ${D}_\textrm{KY}\approx2.4$ for $R=350\ \Omega$), chaoticity was considerably more prominent at $n=1$; further iterations of the fractal apparently introduced chaos-chaos transitions, possibly representing interior crises (Fig. \ref{fig:fig5}a). The frequency spectra became increasingly complex between $n=1$ and $n=3$ (Fig. \ref{fig:fig5}b), closely reflecting the locations of the poles and zeros in the resonators (Fig. \ref{fig:fig1}b and Fig. \ref{fig:fig3}). Elevated geometrical complexity was also apparent on representative Poincar\'{e} sections (Fig. \ref{fig:fig5}c,d), which at $n=1$ illustrated the different situation between hyperchaos at $R=300\ \Omega$ ($\lambda_2\gg0$) and chaos in presence of a broken torus with a complex transverse at $R=1500\ \Omega$ ($\lambda_2\approx0$). Route-to-chaos analysis was beyond the scope of the present numerical analysis, which only had an exploratory purpose, therefore the mechanisms of transition to chaos and hyperchaos need future clarification, particularly in the context of a systematic comparison between simplified and realistic transistor models.\\
As given in Table \ref{tab:tabella3} for two representative settings of the control parameter, i.e. $R=300,1500\ \Omega$, increasing the fractal depth elevated the estimated fractal dimension, reaching ${D}_\textrm{KY}\approx10$ for $n=3$. Albeit distant from filling all available dimensions, i.e. 16 (see Section \ref{oscillator}), this result appears noteworthy in light of the elementary nature of the oscillator circuit, wherein the only non-linearity is that represented by Eq. (\ref{eq:simploscmodel1b}). Over the considered span of fractal depths, the number of positive Lyapunov exponents steadily increased with the fractal iterations, indicating that the system generated high-order hyperchaos (i.e., $\lambda_3>0$). While similar results were previously reported for systems involving delay units, which are asymptotically infinite-dimensional $D\to\infty$ as the quality of the delay approximation is improved, the known transistor-based hyperchaotic oscillators tend to generate considerably lower-dimensional dynamics\cite{elwakil2001,hu2011}. The sum of all Lyapunov exponents was always negative, i.e. $\sum_{i=1}^{4+4n}\lambda_i<0$, confirming that the present system is dissipative\cite{shivamoggi2014}.

\begin{table}
\centering
\caption{Correlation dimension $D_2$ estimated from SPICE simulations of the oscillator containing fractal networks having depth $n=0,1,2$ and realized with identical or mismatched ideal ($Q=\infty$), or realistic ($Q<\infty$) component models. Values given as mean$\pm$standard deviation over the simulation runs.}\label{tab:tabella4}
\begin{tabular}{clcc}
\hline
Depth $n$ & Scenario & $D_2\ (R=300\ \Omega)$ & $D_2\ (R=1500\ \Omega)$\\
\hline
0 & Identical & $1.6$& $2.6$\\
0 & Mismatched & $1.6\pm0.1$ & $2.6\pm0.1$\\
0 & Realistic & $1.3\pm0.3$ & $1.7\pm0.5$\\
1 & Identical & $2.6$ & $3.4$\\
1 & Mismatched & $3.0\pm0.4$ & $3.3\pm1.4$\\
1 & Realistic & $2.7\pm0.3$ & $3.2\pm1.0$\\
2 & Identical & $5.7$ & $5.8$\\
2 & Mismatched & $6.3\pm0.6$ & $7.0\pm1.5$\\
2 & Realistic & $2.2\pm0.7$ & $4.5\pm1.2$\\\hline
\end{tabular}
\end{table}
\subsection{SPICE simulations}\label{oscillator_spice}
To further investigate the system behavior in a more realistic scenario, we next simulated oscillator circuits containing a canonical transistor model and the verbatim fractal networks (Fig. \ref{fig:fig1}a and Fig. \ref{fig:fig4}a), by means of the \verb+ngspice-26+ implementation of the SPICE environment\cite{ngspice}. Namely, we considered the Ebers-Moll model\cite{millman1987} and, reflecting the characteristics of the transistor type chosen for physical realization of the circuits in Section \ref{experimental}, we set transport saturation current $I_\textrm{S}=0.5\textrm{ fA}$, forward and reverse gains $\beta_\textrm{F}=200$, $\beta_\textrm{R}=30$ and base-collector zero-bias depletion capacitance $C_\textrm{JC}=0.2\textrm{ pF}$; essentially equivalent results are obtained with more complex transistor models (data not shown).\\
We performed simulations given $n=0,1,2$ fractal iterations under three separate scenarios. First, in the ``Identical'' scenario all inductances and capacitances were assumed to be exactly homogeneous at each level of the fractal. Reflecting the commercial component values available for circuit realization, the inductances in the first three levels were set to $L_0=22\ \mu\textrm{H}$, $L_1=10\ \mu\textrm{H}$ and $L_2=4.7\ \mu\textrm{H}$, thus deviating slightly from the sequence $L_{i+1}=L_i/2$. Second, in the ``Mismatched'' scenario, the inductances and capacitances, still modeled as ideal components (i.e., quality factor $Q=\infty$) were subject to a random variation within $\pm20\%$ and $\pm5\%$ respectively, which reflects the characteristics of the physical components indicated in Section \ref{experimental}. Third, in the ``Realistic'' scenario, finite quality factor $Q<\infty$ and self-resonance in the individual inductors were introduced, by adding to each one a series resistor and a parallel capacitor, having values $1.7\ \Omega$ and $0.9\textrm{ pF}$, $0.6\ \Omega$ and $1.2\textrm{ pF}$, and $0.3\ \Omega$ and $0.8\textrm{ pF}$ respectively; the equivalent parallel resistors were omitted from these simulations, as the resulting DC paths caused simulation problems (data not shown).\\
Simulations at control parameter values $R=300,1500\ \Omega$ were run until $60\ \mu\textrm{s}$, discarding the first $10\ \mu\textrm{s}$ for initial transient stabilization, and the voltage $v_{C_1}$ (corresponding to $x$ in Eq. \ref{eq:rescoscmodel1}) was stored every $\textrm{d}t=0.5~\textrm{ns}$. For each configuration in the ``Mismatched'' and `Realistic'' scenarios, the simulations were repeated 10 times with different sets of random parametric mismatches.\\
Notably, visual inspection of the time-series reveled a rich repertoire of behaviors, including metastability, intermittency, and situations wherein the random variation of component values substantially affected attractor geometry and even chaoticity. These effects were substantially more prominent compared to the numerical simulations with the simplified model, reported in Subsection \ref{oscillator_numerical}. They are not considered in detail, because SPICE simulations, while important for completeness, should only be treated as indicative given that their reliability in representing the dynamics of transistor-based chaotic circuits is knowingly limited\cite{minati2017}.\\
Since high-dimensional dynamics were expected, the traditional Grassberger-Procaccia algorithm, extensively used in previous studies on chaotic transistor-based oscillators\cite{minati2014,minati2017}, could not be used to estimate the correlation dimension $D_2$ from the time-series, because it is affected by severe issues of underestimation and non-convergence of the correlation sum slopes; preliminary work revealed that for the present signals, these issues could not be sufficiently alleviated solely by improving the time-delay embedding choices\cite{lai1998,kim1999}. We thus tentatively resorted to an improved implementation of the Takens-Ellner algorithm that should enable improved choice of the scaling region of the correlation integral function, and that was previously shown to correctly determine the correlation dimension $D_2\approx8$ of the sum of four Lorenz signals\cite{michalak2011,michalak2014}. The so-called $e$-correction was disregarded, fitting was performed with a polynomial of degree 8, and any negative plateaus were ignored. The complete source code implementing the procedure is available from Ref. \citen{hdstoolkit}.\\
As given in Table \ref{tab:tabella3}, across the scenarios and settings of the control parameter $R$, the correlation dimension $D_2$ generally increased with the fractal depth $n$, confirming the relationship initially revealed by the simplified model. For $n>0$, we also observed that generally $D_2<D_\textrm{KY}$: besides the changed transistor model alongside possible estimation and numerical aspects, this plausibly reflects fundamental differences in the measures themselves, particularly in that the former reflects topology whereas the latter is based on dynamics\cite{chlouverakis2005}. Of relevance to the experimental realizations described next, comparison of the three scenarios further suggested that the generation of high-dimensional dynamics was relatively insensitive to the mismatches introduced by component value tolerances, however it could be appreciably hindered by non-ideal inductor behavior already at $n=2$. This situation, wherein component non-idealities seemingly have a greater impact on the dynamics than parametric tolerances, resembles some earlier observations in single-transistor chaotic oscillators without fractal elements\cite{minati2014}.\\

\begin{table*}
\centering
\caption{Correlation dimension and permutation entropy estimated, at the given settings of $R$ and $C_1$, from experimental measurements of fractal networks having depth $n=0,1,2$ (Fig. \ref{fig:fig1}a and Figs. \ref{fig:fig6}a,c), including fractals having randomly reshuffled inductor values (Figs. \ref{fig:fig8}a,b), imperfect fractals (Figs. \ref{fig:fig8}c,d) and Foster equivalent networks (Fig. \ref{fig:fig1}c and Figs. \ref{fig:fig6}b,d). $\langle D_2\rangle$ and $\langle h_5\rangle$ respectively denote the average correlation dimension and order-5 permutation entropy for the measured time-series; $\langle \hat{D}_2\rangle$ and $\langle\hat{h}_5\rangle$ refer to the corresponding surrogates, $\langle l\rangle$ represents the length of the map-like time-series for permutation entropy calculation. Values averaged over circuit board specimens and settings of $R$, and given as mean$\pm$standard deviation. Full set of measurements available in Supplementary Table S2.}\label{tab:tabella5}
\begin{tabular}{clccccccc}
\hline
Depth $n$ & Configuration & $\langle R\rangle$ & $\langle C_1\rangle$ & $\langle D_2\rangle$ & $\langle \hat{D}_2\rangle$ & $\langle l\rangle$ & $\langle h_5\rangle$ & $\langle\hat{h}_5\rangle$\\
\hline
0 & Regular fractal & $327\pm9\ \Omega$ & $295\pm14\textrm{ pF}$ & $1.1\pm0.1$ & $3.3\pm0.1$ & $4669\pm694$ & $0.29\pm0.03$ & $0.63\pm0.01$\\
1 & Regular fractal & $163\pm6\ \Omega$ & $288\pm6\textrm{ pF}$ & $2.0\pm0.1$ & $4.1\pm0.1$ & $5578\pm29$ & $0.45\pm0.02$ & $0.61\pm0.01$\\
2 & Regular fractal & $805\pm679\ \Omega$ & $290\pm5\textrm{ pF}$ & $2.7\pm0.1$ & $6.6\pm0.4$ & $3913\pm301$ & $0.55\pm0.03$ & $0.69\pm0.01$\\
2 & Equivalent network & $113\pm51\ \Omega$ & $285\pm6\textrm{ pF}$ & $3.5\pm0.4$ & $6.2\pm0.7$ & $11343\pm257$ & $0.63\pm0.02$ & $0.69\pm0.01$\\
2 & Randomly reshuffled fractal & $1812\pm234\ \Omega$ & $288\pm1\textrm{ pF}$ & $3.8\pm0.2$ & $5.6\pm0.3$ & $4020\pm126$ & $0.63\pm0.01$ & $0.69\pm0.00$\\
2 & Imperfect fractal & $510\pm389\ \Omega$ & $280\pm2\textrm{ pF}$ & $4.2\pm0.7$ & $6.4\pm0.6$ & $3115\pm329$ & $0.61\pm0.04$ & $0.71\pm0.01$\\
\hline
\end{tabular}
\end{table*}
\section{EXPERIMENTAL REALIZATIONS}\label{experimental}
\begin{figure*}
\centering
\includegraphics[width=1\textwidth]{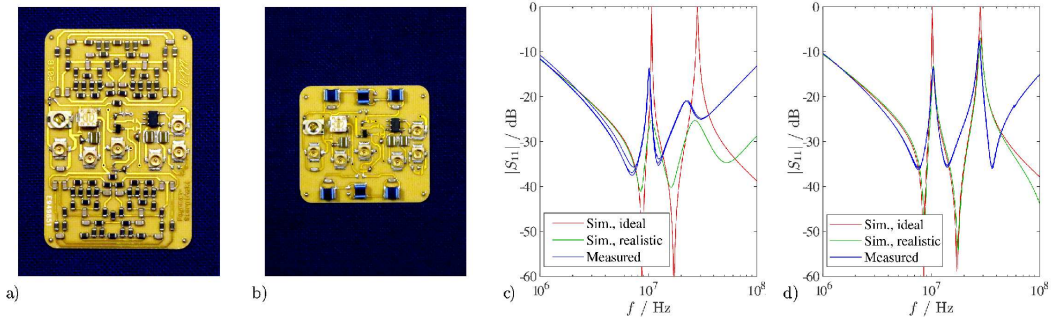}
\cprotect\caption{Physical construction of the oscillator containing the fractal resonators. a) Realized board (size $28\times40\textrm{ mm}$) containing two verbatim instances (top, bottom) of the fractal with $n=2$, as per Fig. \ref{fig:fig1}a. b) Realized board (size $28\times25\textrm{ mm}$) containing two instances (top, bottom) of the corresponding Foster equivalent network, as per Fig. \ref{fig:fig1}c. c) Simulated and experimentally-measured frequency responses for the verbatim realization. d) Simulated and experimentally-measured frequency responses for the equivalent network realization. The simulated responses are shown separately for ideal ($Q=\infty$) and realistic ($Q<\infty$, component manufacturer models) cases, and the experimental responses were measured separately for the three circuit specimens.\label{fig:fig6}}
\end{figure*}

\begin{figure*} 
\centering
\includegraphics[width=1\textwidth]{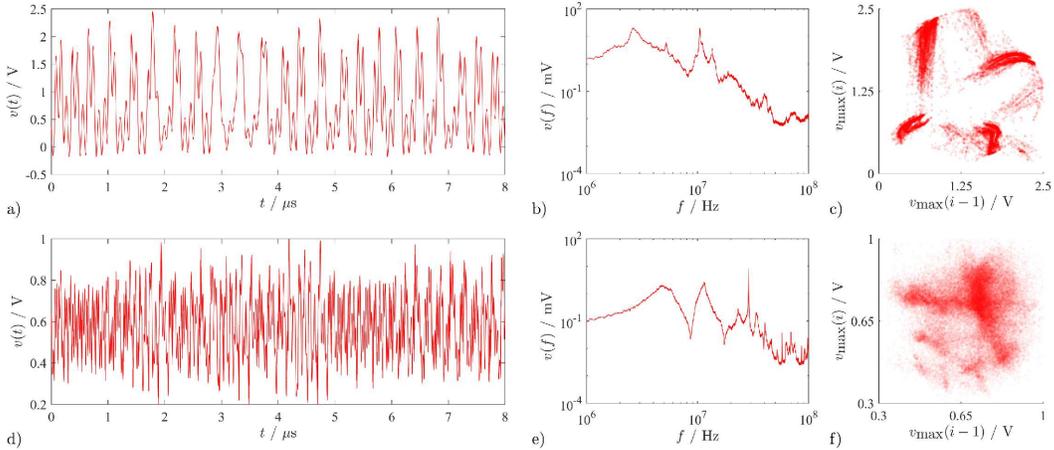} 
\cprotect\caption{Representative experimental data from realizations of the regular fractal having depth $n=2$. a) Time-series, b) Frequency spectrum, smoothed for visualization, and c) Zero temporal-derivative Poincar\'{e} section at $R=186\ \Omega$ for an oscillator containing verbatim instances of the fractal as in Fig. \ref{fig:fig1}a and Figs. \ref{fig:fig6}a,c (dataset $i=11$ in Supplementary Table S2, $D_2=2.6\pm0.1$ and $h_5=0.53\pm0.01$). d) Time-series, e) Frequency spectrum, smoothed for visualization, and f) Zero temporal-derivative Poincar\'{e} section at $R=174\ \Omega$ for an oscillator containing the Foster equivalent implementation of the fractal as in Fig. \ref{fig:fig1}c and Figs. \ref{fig:fig6}b,d (dataset $i=40$ in Supplementary Table S2, $D_2=4.1\pm0.3$ and $h_5=0.66\pm0.00$).\label{fig:fig7}}
\end{figure*}

\subsection{Physical construction and setup}\label{experimental1}
The fractal resonator at iterations $n=0,1,2$ was physically realized verbatim by means of discrete components, and inserted twice in the initial oscillator circuit (Fig. \ref{fig:fig6}a); a single board design was prepared, and the cases $n<2$ were obtained by depopulating the corresponding components. As detailed in Supplementary Table S1, commercially-available miniaturized inductors and capacitors in standard imperial format 0603 (size $1.6\times0.8\textrm{ mm}$) were chosen for realizing the two large LC networks while minimizing parasitics and mismatches. The corresponding SPICE models of all inductors are available from the respective manufacturers\cite{indmodels}; the non-idealities of the capacitors are assumed to be negligible.\\
Similarly to Ref. \citen{minati2017}, a ceramic substrate was chosen for circuit realization, and the NPN bipolar junction transistor was of type PRF949 (NXP Semiconductor, Eindhoven, The Netherlands). Voltage $v_\textrm{C1}$ was selected as the physical variable of interest and, to minimize loading effects, a low-capacitance preamplifier of type MAX4200 (Maxim Inc., San Jose CA, USA) was provisioned. A capacitive trimmer was also installed to allow adjusting $C_1$ in order to avoid the onset of spurious high-frequency oscillations unrelated to the resonator circuits. As in Ref. \citen{minati2017}, the resistor $R$ was implemented as a trimmer, and manually adjusted to explore different regions of chaotic operation for each circuit. The resulting circuit board was equipped with six U.FL coaxial sockets, allowing connection to the external power supply for the oscillator ($V_\textrm{CC}=+2.5\textrm{ V}$) and the preamplifier ($V_\textrm{DD}=+5\textrm{ V}$, $V_\textrm{EE}=-1\textrm{ V}$), and connection to an oscilloscope for time-series digitization or to a network analyzer for frequency response measurement.\\
A network equivalent to the fractal at iteration $n=2$ (Fig. \ref{fig:fig1}c) was also physically realized and inserted in the oscillator circuit (Fig. \ref{fig:fig6}b). Given the substantially reduced number of components, inductors in the larger standard imperial format 1008 (size $2.5\times2.0\textrm{ mm}$) could be used; owing to the lower inductance values needed and the different construction, these had superior frequency characteristics. In this case, the inductors were not selected based on the Foster method, but considering the available types from a commercial catalog, aiming to realize as closely as possible the response of the fractal while taking into account the non-ideal features of the physical components. The resulting network is a Foster network with three series blocks, each consisting of a parallel inductor and capacitor: the first having only an inductor of value $3.3\ \mu\textrm{H}$, the second with values $4.7\ \mu\textrm{H}$ and $18\textrm{ pF}$, and the third with values $4.7\ \mu\textrm{H}$ and $75\textrm{ pF}$ (details in Supplementary Table S1).\\
To evaluate reproducibility, three specimens were realized for each configuration. The circuits were supplied by a low-noise power supply type 6627A (Keysight Technologies, Inc., Santa Rosa CA, USA). The time-series were digitized into 4,000,000 points at 8-bit resolution, 2 GSa/s sampling rate in DC-coupled, $50\ \Omega$ terminated configuration,  using a digital-storage oscilloscope type WaveJet 354T (LeCroy Inc., Chestnut Ridge NY, USA). The frequency responses of the resonators were measured using a digital $S$-parameter vector network analyzer type 8753ES (Keysight Technologies, Inc.) connected in two-port configuration and fully calibrated using a microwave load resistor on the oscillator circuit board itself. All board design materials, raw time-series and frequency response datasets have been made freely available\cite{onlinemat}.
\subsection{Data analysis}\label{experimental2}
In order to attenuate the discretization and digitization noise, each time-series was smoothed with a moving average filter spanning 15 samples. For determining the correlation dimension $D_2$, the procedure described in Section \ref{oscillator_spice} was applied over 10 evenly-spaced segments of 100,000 points, which were extracted from the recorded signals. Furthermore, to illustrate the convergence of the dimension determination, for each segment the corresponding surrogate was obtained through a procedure which preserves the Fourier amplitudes and value distribution while destroying all non-linear correlations\cite{schreiber2000}. Each segment was compared to the corresponding surrogate by means of a paired $t$-test; the analyses of each recorded segment and of its surrogate were completely identical and independent.\\
Ever though its detailed assessment is beyond the scope of this work, we furthermore searched for possible signatures of multifractality using the same detrended fluctuation analysis considered in Ref. \citen{minati2017}. As in that study, no convincing evidence was obtained, and consideration of the Fourier surrogates highlighted spurious effects caused by oscillations at multiple scales. For brevity, these results are  omitted.\\ 
Because estimating the fractal dimension of high-dimensional dynamics is knowingly vulnerable to methodological and numerical pitfalls, we confirmed the effect of the fractal depth $n$ and the other manipulations described in Subsections \ref{experimental3} and \ref{experimental4} also by means of permutation entropy. This highly robust measure of complexity is based on the relative frequencies of the possible patterns of symbol sequences\cite{bandt2002}, defined on the basis of the relative ranks of samples $x_i,x_{i+\tau},\ldots,x_{i+(m-1)\tau}$ in a time-series $X=\{ x_i\}$ for $i=1,\ldots,N$. For this analysis, the time-series were converted to a map-like representation by extracting the sequence of alternating local maxima and minima, hence we set $\tau=1$. As advocated in recent guidelines, we furthermore set order $m=5$, which represents the highest setting for which $5m!<N$ for all time-series under consideration, and therefore refer to the permutation entropy with $h_5$\cite{riedl2013}. Comparison to surrogate data was performed as indicated for $D_2$.\\
Across all realized circuit board specimens and parameter regions identified by tuning $R$, a total of 51 time-series were acquired. The corresponding characteristics are reported in full in Supplementary Table S2. Remarkably, a strong correlation was observed between $D_2$ and $h_5$, with rank-order $\rho=0.77$, $p<0.001$. This result provides reassurance that, despite the different underpinnings, both measures indexed the complexity of the chaotic dynamics; nevertheless, consideration of the scatter-plot (not shown) indicated that $h_5$ tended to saturate for high dimensions, i.e. $D_2>3$.
\subsection{Regular fractals and equivalent networks}\label{experimental3}
The frequency response of the fractal resonator realized verbatim at depth $n=2$ was calculated assuming either ideal ($Q=\infty$) or realistic ($Q<\infty$, parallel capacitance and resistance, series resistance) component models, in both cases neglecting the parametric mismatches (tolerances). As expected, the measured responses of the physically-realized resonators were markedly closer to the realistic simulations. In contrast with the ideal scenario, they featured shallow resonances and non-negligible power dissipation, i.e. the poles and zeros were not purely imaginary, even though their frequency locations were approximately preserved (Fig. \ref{fig:fig6}c).\\
As regards the complexity of chaotic dynamics in the oscillators, for succinctness only the averages computed over all circuit board specimens and settings of $R$ are provided in Table \ref{tab:tabella5}. Overall, the relationship between the fractal depth $n=0,1,2$ and attractor dimension observed in the numerical simulations with the simplified model (Subsection \ref{oscillator_numerical}) and in the SPICE simulations (Subsection \ref{oscillator_spice}) was confirmed by the experiments, with corresponding correlation dimension $\langle D_2\rangle=1.1,2.0,2.7$ and order-5 permutation entropy $\langle h_5\rangle=0.29,0.45,0.55$ (rank-order $\rho=0.80$, $p<0.001$ for both measures).\\
In nearly all cases, both measures were significantly lower for the experimental data than the corresponding surrogates ($t$-test $p<0.005$), corroborating the validity of the estimations. The location of the chaotic regions was, however, only weakly related to the bifurcation diagrams determined numerically via the simplified model (Fig. \ref{fig:fig5}a). The reproducibility across circuit board specimens and control parameter settings was good, with a coefficient of variation of 5.5\% for both $\langle D_2\rangle$ and $\langle h_5\rangle$ at depth $n=2$. Intermittency occurred more rarely compared to the SPICE simulations, metastability was not evident, and chaotic dynamics were absent only in the oscillators including the fractals of depth $n=0$.\\
A time-series representative of the experimental circuit containing resonators with $n=2$ demonstrates oscillations at multiple temporal scales, consisting of small-amplitude, fast fluctuations overlapped to larger-amplitude, slower dynamics (Fig. \ref{fig:fig7}a). In agreement with the simulations, the frequency spectrum of activity appreciably reflected the response of the fractal resonator elements (Fig. \ref{fig:fig7}b vs. Fig. \ref{fig:fig6}c). The resulting attractor, visualized as a Poincar\'{e} section, appeared of intermediate complexity and had a characteristic bun-like shape (Fig. \ref{fig:fig7}c).\\
Compared to the verbatim realization, the simulated frequency responses of the equivalent network featured improved similarity between the ideal and the realistic scenarios, indicating that this form of realization was not only markedly more compact (5 vs. 51 components in a resonator), but also electrically superior. Accordingly, agreement with the experimental measurements was also enhanced, even though higher losses with respect to simulation remained evident (Fig. \ref{fig:fig6}d).\\
Compared to the verbatim realization, significantly more complex dynamics were obtained for this implementation, according to both the correlation dimension (i.e., $\langle D_2\rangle=3.5\pm0.4$ vs. $2.7\pm0.1$, $p<0.001$) and the order-5 permutation entropy (i.e., $\langle h_5\rangle=0.63\pm0.02$ vs. $0.55\pm0.03$, $p<0.001$).\\
Plausibly owing to the fact that the higher-frequency poles and zeros were more prominent, faster fluctuations were evident in the time-series (Fig. \ref{fig:fig7}d) and in the associated frequency spectrum (Fig. \ref{fig:fig7}e), which even more closely reflected the response of the resonator (Fig. \ref{fig:fig6}d). The Poincar\'{e} section was markedly different, in that the bun-like shape was replaced by a point cloud with less clear structure, plausibly as a consequence of the higher-dimensional dynamics (Fig. \ref{fig:fig7}f).\\\\

\begin{figure*}
\centering
\includegraphics[width=1\textwidth]{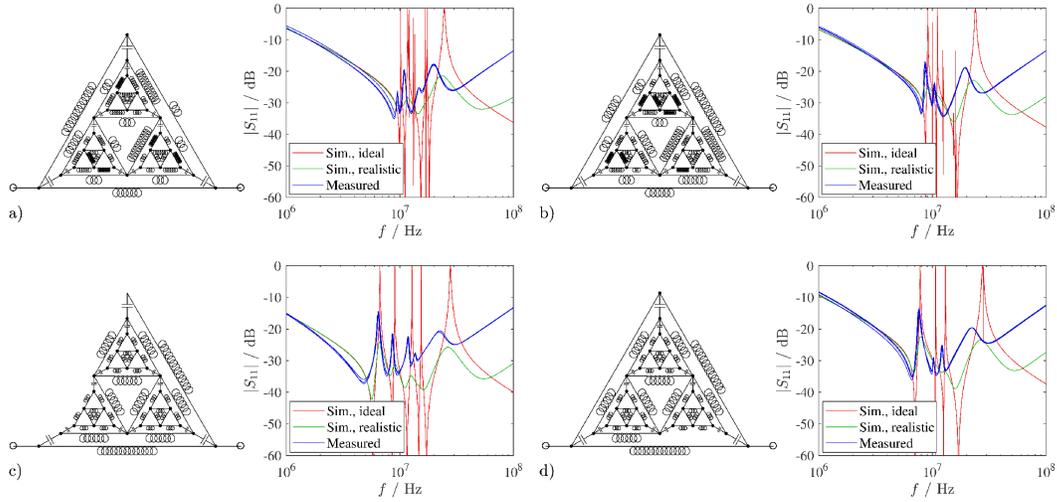} 
\cprotect\caption{Irregular fractals and corresponding frequency responses. a) and b) Networks with intact fractal topology but reshuffled inductor values. c) and d) Fractals rendered imperfect by removal or shorting of two inductors. The networks in a) and c) were inserted to realize $Z_1$, the networks in b) and d) were inserted to realize $Z_2$. Simulated responses are shown separately for ideal ($Q=\infty$) and realistic ($Q<\infty$, component manufacturer models) networks, and experimental responses were measured separately for the three circuit specimens. Number of turns in inductor symbols is proportional to inductance.\label{fig:fig8}}
\end{figure*}

\begin{figure*}
\centering
\includegraphics[width=1\textwidth]{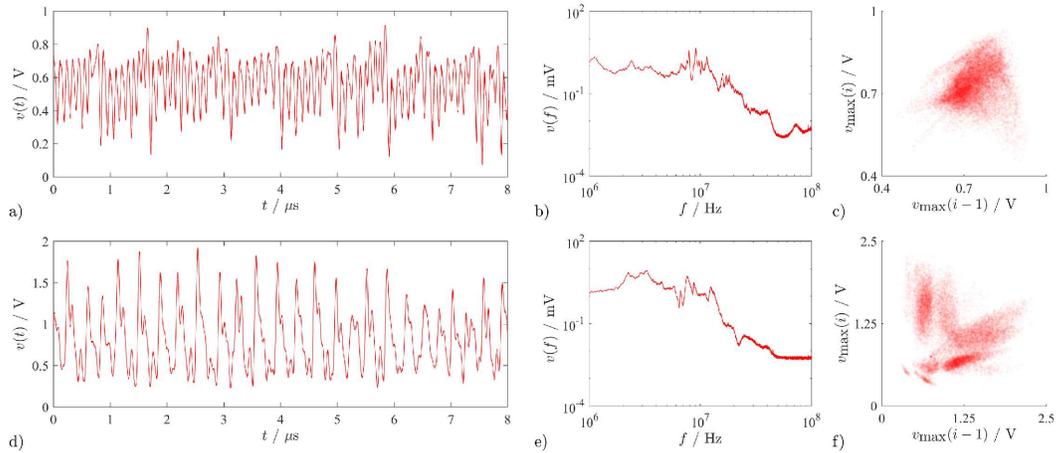} 
\cprotect\caption{Representative experimental data from realizations of the irregular fractals. a) Time-series, b) Frequency spectrum, smoothed for visualization, and c) Zero temporal-derivative Poincar\'{e} section at $R=2033\ \Omega$ for an oscillator containing reshuffled inductor values as in Figs. \ref{fig:fig8}a,b (dataset $i=26$ in Supplementary Table S2, $D_2=4.1\pm0.2$ and $h_5=0.64\pm0.00$). d) Time-series, e) Frequency spectrum, smoothed for visualization, and f) Zero temporal-derivative Poincar\'{e} section at $R=56\ \Omega$ for an oscillator containing imperfections as in Figs. \ref{fig:fig8}c,d (dataset $i=39$ in Supplementary Table S2, $D_2=4.9\pm0.5$ and $h_5=0.61\pm0.01$).\label{fig:fig9}}
\end{figure*}
\subsection{Irregular fractals}\label{experimental4}
The fractal structures found in nature are not only truncated, but also highly imperfect\cite{mandelbrot1983,mandelbrot2004,havlin1995}. This is well-evident for dendritic trees and axonal branches which, while having fractal and possibly even multifractal topology at a statistical level, in practice pervasively deviate from it\cite{caserta1990,alves1996}. In particular, it appears noteworthy to consider two distinct aspects. First, even in presence of repeated branching which gives rise to self-similar structure, fundamental physiological properties may not at all scale accordingly. For example, the intensity of synapses is strongly influenced by Hebbian learning, and not directly related to the point at which they are found in a dendritic tree\cite{brito2016}. Second, frequent topological imperfections are found, for example in the form of missing or irregular branching. Inspired by these observations, we attempted to replicate similar features in the present fractal resonators, to investigate their effect on frequency response and oscillatory dynamics\cite{werner2010,paar2001}.\\
In order to do so, we initially randomly reshuffled the inductor values across levels of the fractal having depth $n=2$, additionally altering in an arbitrary manner their relative proportions. The corresponding LC networks were physically realized verbatim. Even though the discrepancy between the simulated and measured frequency responses was greater for these irregular networks, in all cases it was well-evident that vastly more complex resonances, with up to 5 pairs of visible poles and zeros, were obtained compared to the regular case (Fig. \ref{fig:fig8}a,b).\\
Accordingly, significantly richer dynamics were obtained by instancing the two different resonators as $Z_1$ and $Z_2$ in the oscillator circuit. Compared to the oscillator containing the regular realization of the fractal, dynamics were more complex according to both the correlation dimension (i.e., $\langle D_2\rangle=3.8\pm0.2$ vs. $2.7\pm0.1$, $p<0.001$) and the order-5 permutation entropy (i.e., $\langle h_5\rangle=0.63\pm0.01$ vs. $0.55\pm0.03$, $p<0.001$). In this case, the illustrative time-series was characterized by cycle amplitude fluctuations having a stronger periodic component but highly irregular amplitude fluctuations (Fig. \ref{fig:fig9}a). This was reflected in a flatter frequency spectrum (Fig. \ref{fig:fig9}b), wherein there was no evident correspondence with the poles and zeros, plausibly because of the considerably more complex resonances and mismatch between the two branches. There was virtually no discernible structure on the Poincar\'{e} section (Fig. \ref{fig:fig9}c).\\
We next considered the more parsimonious case of focal imperfections, which were introduced in the form of up to two missing or short-circuited inductors in each resonator. Strikingly, this operation realized resonances having a level of complexity similar to or even higher than those obtained by the much more demanding reshuffling operation (Fig. \ref{fig:fig8}c,d).\\
Accordingly, also in this case significantly richer dynamics were obtained by instancing the resonators in the oscillator circuit, as indicated by both the correlation dimension (i.e., $\langle D_2\rangle=4.2\pm0.7$ vs. $2.7\pm0.1$, $p<0.001$) and the order-5 permutation entropy (i.e., $\langle h_5\rangle=0.61\pm0.04$ vs. $0.55\pm0.03$, $p<0.001$). There were, incidentally, no significant differences ($p=0.1$) for either measure with respect to the randomly reshuffled realization. More similarly to the regular fractal case, in this case the illustrative time-series was again characterized by overlapped large, slow fluctuations and smaller, faster fluctuations (Fig. \ref{fig:fig9}d). However, the frequency spectrum remained relatively flat (Fig. \ref{fig:fig9}e), without evident correspondence to the resonances. The Poincar\'{e} section had the appearance of multiple, partially distinct point clouds (Fig. \ref{fig:fig9}f).\\
\section{DISCUSSION}\label{discussion}
Although elevated system order is not sufficient to ensure the onset of high-dimensional chaos, it is a prerequisite. Many hyperchaotic circuits have been obtained empirically, by increasing system dimensionality through complicating a preexisting oscillator, and then exploring the parameter space searching for regions yielding two or more positive Lyapunov exponents. This approach is exemplified by adding time-delayed state feedback to the Chen system, which yields hyperchaotic dynamics with up to 11 (in numerical simulations) or 5 (in experiments) positive Lyapunov exponents\cite{hu2011}. In general, time-delay systems, which are intrinsically infinite-dimensional, often enable obtaining high-dimensional chaos with relatively simple circuits\cite{namajunas1995,buscarino2011}. The time-delay block has frequently been implemented in the form of a chain of either active\cite{hu2011} or passive\cite{namajunas1995} filters, wherein the approximation of an ideal time-delay element becomes increasingly accurate as the number of stages is elevated. This approach effectively yields a dimensionality which increases with the accuracy of the approximation. Despite the different underlining framework, such relationship matches our findings on the effect of the number of fractal iterations (depth) on the correlation dimension and number of positive Lyapunov exponents.\\
Another empirical approach to the design of hyperchaotic circuits involves increasing system dimensionality by coupling two (or more) low-order circuits in such a manner as to hinder their synchronization: towards this aim, one may implement either weak linear (e.g., resistive) coupling\cite{vcenys2003,kapitaniak1994}, or nonlinear coupling\cite{murali2000,lindberg2001,cannas2002}. Incidentally, this literature also motivates investigating the synchronization behavior of the oscillators introduced in this study.\\
On the other hand, some hyperchaotic circuits have been obtained through applying first principles of non-linear dynamics; as such, they involve particularly simple topologies, and accordingly a limited number of components\cite{tamasevicius1996,tamasevicuius1997,elwakil1999,li2014}. In this regard, it appears noteworthy that, at least in numerical simulations based on the simplified model, the present circuit appears to generate in limited cases high-dimensional chaos already when including Feynman-Sierpi\'{n}ski resonator of depth 1. Deeper fractals require considerably more components for verbatim realization, however, as shown in Section \ref{fractal}, the impedance of regular fractals can be conveniently realized in the form of chain networks of parallel inductors and capacitors: for the case of a fractal of depth 2, the total resulting number of components in the oscillator is 13 (Fig. \ref{fig:fig6}b). The circuit size is therefore relatively low compared to the known designs of hyperchaotic oscillators: in particular, it appears noteworthy that complex components such as multipliers are absent and the only active, non-linear component in the present circuit is a transistor.\\
Insofar as some meaningful analogy can be postulated between the behavior of non-linear electronic circuits and that of neurons\cite{minati2014,minati2017}, two remarks of potential neuroscientific relevance can be made. First, the depth of truncated fractals plausibly impacts dynamical complexity in a general sense. This relationship was first put forward in the context of truncated fractal patterns generated by coupled non-linear oscillators\cite{paar2001}. Here, by exploring the effect of fractal depth in a more strictly structural sense, it was confirmed that deeper self-similar morphology engenders richer dynamics. This leads to the testable hypothesis that the level of branching observed for neurites, such as dendritic trees and axonal processes, is related to dynamical complexity as would be reflected, for example, in the temporal irregularity of action potential generation. A similar relationship could exist at the meso- and macroscopic scales, with reference to the layout of axonal bundles. Second, irregularities and imperfections enhance the dynamical complexity by corrupting the self-similarity and introducing additional resonances. Such situation is directly reminiscent of well-known observations from materials science, wherein impurities and defects can produce a plethora of intricate spectra, and may strongly impact the bulk properties of media\cite{bridges1990,stillinger1995,raebiger2010}. In turn, this leads to hypothesizing that, in neurites, deviations from perfect regularity may not just arise out of biological difficulty in realizing perfect fractals, but perhaps also have a functional significance.
\section*{SUPPLEMENTARY MATERIAL}
See supplementary material for components chosen for physical realization of the oscillators, and dynamical parameters for all measured time-series.
\section*{ACKNOWLEDGMENTS}
All experimental activities were self-funded by L.M. personally and conducted on own premises. The authors are sincerely grateful to Luca Faes (University of Palermo, Italy) for fruitful discussions about information content estimation, and to Krzysztof Michalak (University of Pozna\'{n}, Poland) for making the HDS toolkit publicly available and for providing outstanding support and advice on its usage.

\end{document}